\documentclass[a4paper,11pt]{article}
\usepackage{pos}

\usepackage{graphicx}
\usepackage{amssymb}
\usepackage{amsmath,esint}
\usepackage{lineno}
\usepackage{subfig}
\usepackage{setspace}

\title{A calibration study of local ice and optical sensor properties in IceCube}
 \ShortTitle{A calibration study of local ice and optical sensor properties in IceCube}

\author{The IceCube Collaboration \\{\normalsize \normalfont(a complete list of authors can be found at the end of the proceedings)}}

\emailAdd{dima@icecube.wisc.edu}

\abstract{The optical sensors of the IceCube Neutrino Observatory are attached on vertical strings of cables. They were frozen into the ice in the deployment holes made by hot water drill. This hole ice, to the best of our knowledge, consists of a bubbly central column, with the remainder of the re-frozen volume being optically clear. The bubbly ice often blocks one or several of the calibration LEDs in every optical sensor and significantly distorts the angular profile of the calibration light pulses. It also affects the sensors’ response to in-coming photons at different locations and directions. We present our modeling of the hole ice optical properties as well as optical sensor location and orientation within the hole ice. The shadowing effects of cable string and possible optical sensor tilt away from the nominal vertical alignment are also discussed.

\vspace{4mm}
{\bfseries Corresponding author:}
Dmitry Chirkin$^{1*}$\\
{$^{1}$ Dept. of Physics and Wisconsin IceCube Particle Astrophysics Center,\\ University of Wisconsin, Madison, WI 53706, USA}\\[4mm]
$^*$ Presenter

\affiliation[a]{}

}

\FullConference{37$^{\rm{th}}$ International Cosmic Ray Conference (ICRC 2021)\\
		July 12th -- 23rd, 2021\\
		Online -- Berlin, Germany}

\begin{document}
\maketitle

\section{Introduction}

The IceCube Neutrino Observatory is a cubic-kilometer neutrino  detector installed in the ice at the geographic South Pole \cite{detector} instrumenting depths between 1450\,m and 2450\,m. It detects Cherenkov radiation emitted by secondary particles produced in neutrino interactions in the surrounding ice or the nearby bedrock. The optical  properties of ice surrounding the detector have been studied in detail, and the latest results are presented at this conference \cite{bfr}. This paper focuses on the properties of ice in the immediate vicinity of the optical sensors (digital optical module, or DOM), and on the relative position and orientation of the DOMs within this local ice.

Each of the 5160 DOMs deployed in ice is equipped with 12 light emitting diodes (LEDs) that are positioned on a daughter "flasher" board and can emit light one at a time or in simultaneous combinations. The LEDs are placed in pairs (1 and 7, 2 and 8, and so on) at 60-degree increments in azimuth. LEDs 1-6 emit light at an angle of 48 degrees up, and LEDs 7-12 point straight out horizontally. Most of the instrumented LEDs emit light centered at 405 nm wavelength in a cone of about 9.7 degree width (RMS).

Initial {\it all-purpose} calibration sets (taken mainly to study the optical properties of ice) used configuration in which all 6 horizontal LEDs were emitting light simultaneously, creating an approximately azimuthally-symmetric pattern of light. This light was observed by the DOMs on the surrounding strings and the resulting data sets were used to fit the properties of inter-string ice. Because of the abundance of data (high number of emitter-receiver pairs) the simple hypothesis of cylindrically-symmetric angular profile of the emitted light was sufficient to achieve good description of flasher data and to extract the ice parameters.

To improve precision of the fits we experimented with an unfolding procedure that fitted the azimuthal profile of the emitted light for each of the emitting DOMs individually. An array of 72 directions spaced by 5 degrees in azimuth (covering the entire 360 degree range), together with one up and one down components (motivated by simulation of possible unintended up/down scattering of light at the point where it entered the ice), was simulated and the best linear combination was found for each tested hypothesis during fitting of the ice properties. Unsurprisingly, this did lead to an improvement in the quality of flasher data description with the fit. When we used this unfolding procedure on data with only one LED flashing at a time, it did in fact produce a shape in the unfolded components that peaked around a single azimuthal direction. Such directions were spaced out by approximately 60 degrees on a circle, corresponding to pointing directions of individual LEDs. However, this method was ultimately abandoned once we noticed that the combined histogram of all components of all horizontal LEDs on all DOMs in the detector did not form a rectangular distribution but rather an oscillating shape, which aligned with the direction of the ice flow. This problem was eventually solved with the new birefringence-based description of ice anisotropy (described in \cite{bfr}), but by then we switched to a richer single-LED flasher data set.

\section{Fits to DOM azimuthal orientation and cable position}
\begin{figure}[h]
    \centering
    \includegraphics[width=0.32\textwidth]{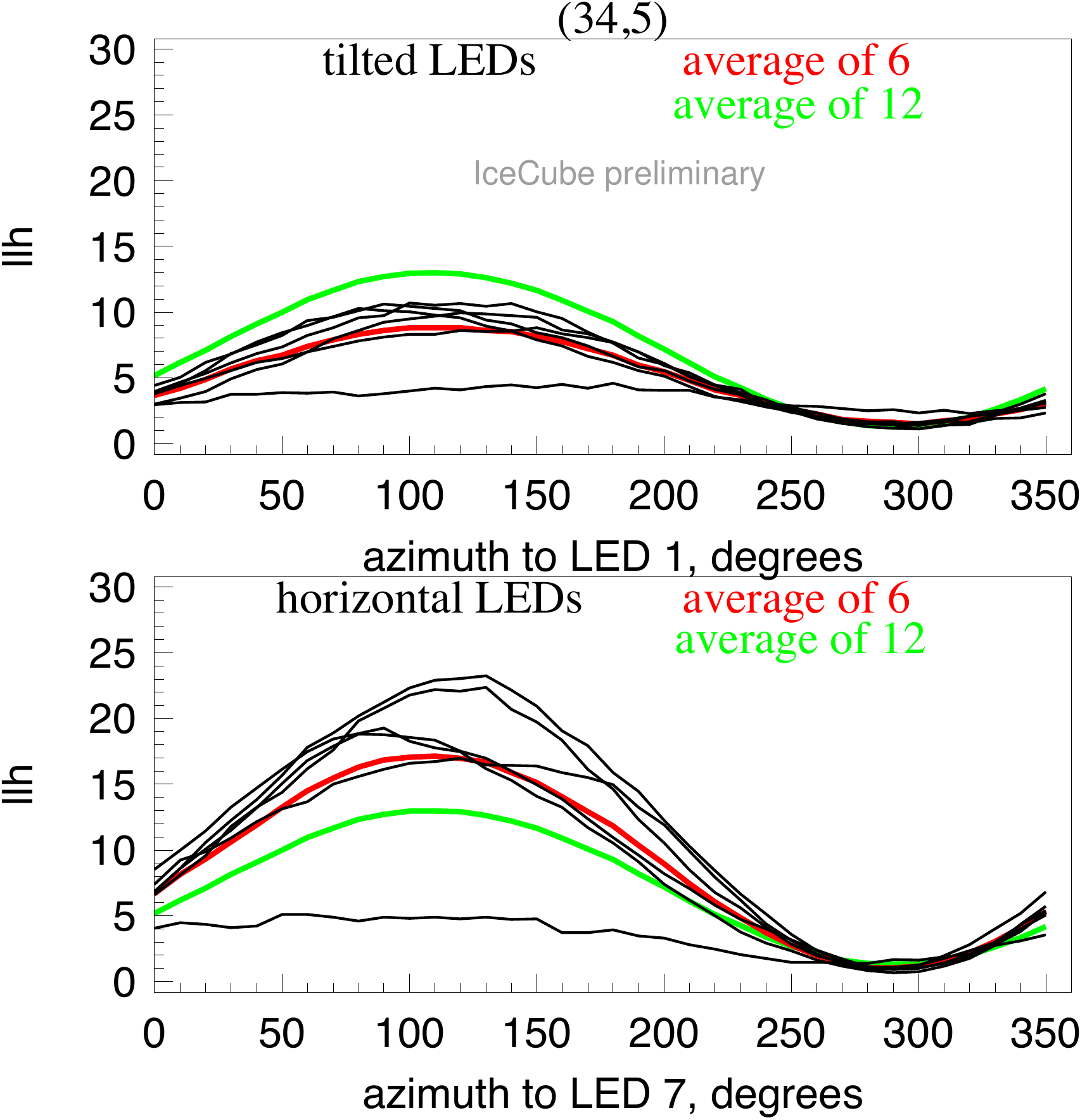}\hspace{1cm}
    \includegraphics[width=0.32\textwidth]{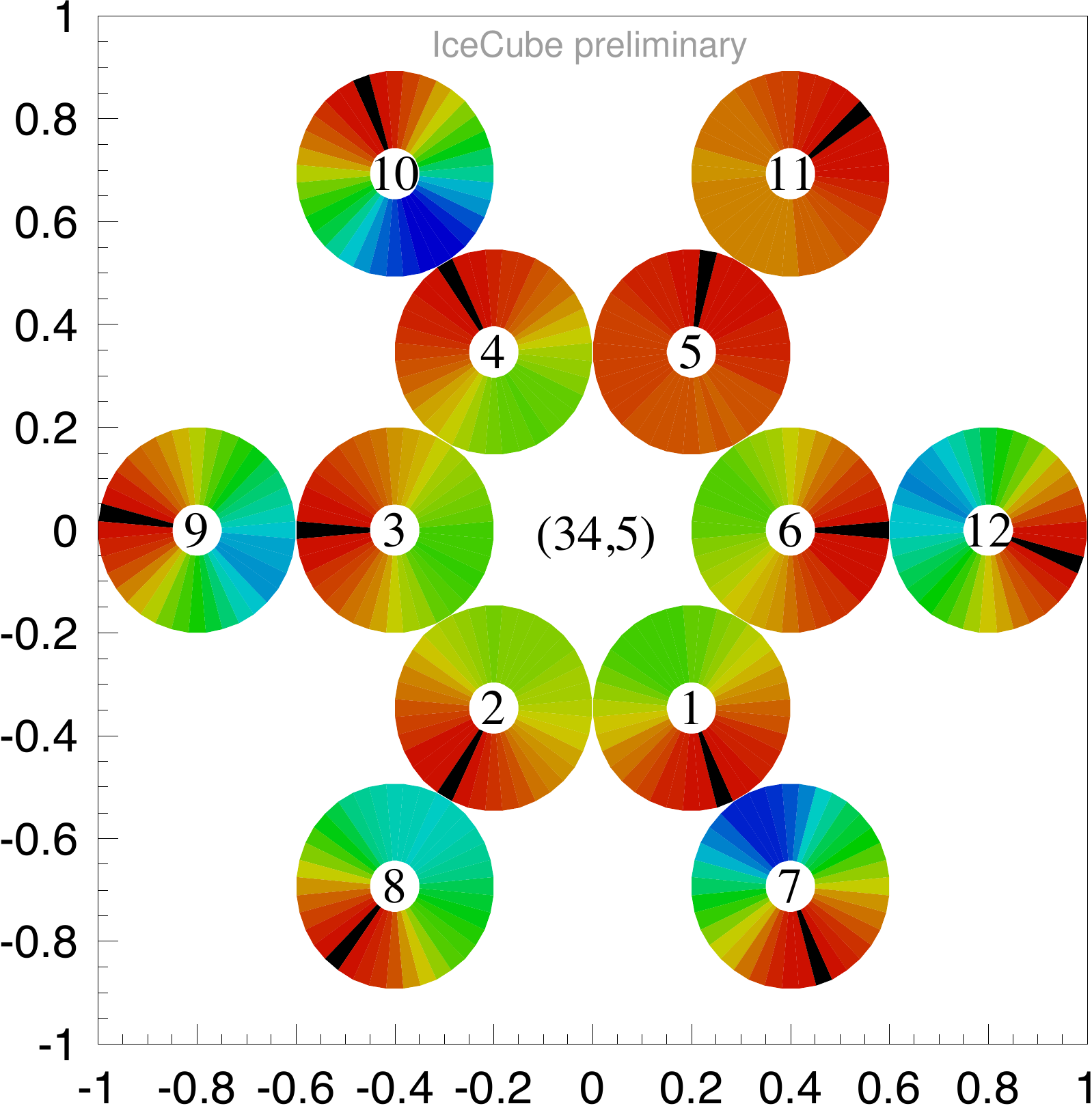}
    \includegraphics[width=0.0125\textwidth]{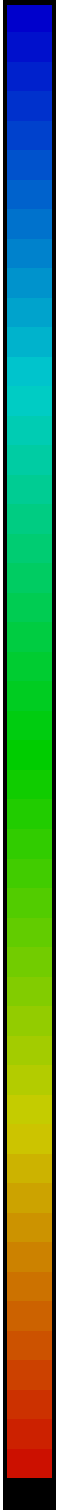}
    \caption{Left: minus log likelihood ({\it llh} - see the last section for details) curves vs.\ azimuth for LEDs 1-6 (top) and LEDs 7-12 (bottom). Curves shifted by multiples of 60 degrees as appropriate to align with each other. Right: same is shown with color scale (from blue - worst {\it llh} to red/black - best {\it llh}), LED circles placed in their reconstructed azimuthal positions. Best {\it llh} values are shown with black in each circle and correspond to reconstructed directions of LED "beams". The span of the color scale is the same in all circles. LEDs 5 and 11 stand out with a very flat {\it llh} landscape.}
    \label{azimuth}
    \vspace{-0.2cm}
\end{figure}

The {\it single-LED} data set had LEDs 1-12 (i.e. all of tilted and horizontal) flashing one at a time on all working in-ice DOMs. Since we could no longer approximate such data with a cylindrically-symmetric pattern, a simulation campaign was carried out in which azimuthal orientations of all DOMs in the detector were determined. For each LED, we simulated precise (lab-measured) angular profile of the LED in ten-degree increments of azimuth angle, with the nominal tilt (of 48 degrees up for tilted and 0 for horizontal LEDs). An example of the resulting likelihood curves is shown in figure \ref{azimuth}. With usually no more than two exceptions per DOM, all curves align with their peaks and valleys, when shifted by multiples of 60 degrees (as required to match the nominal positions of LEDs on the flasher board). We have therefore combined all 12 such curves (per DOM) and found that the combined curve can be very well fitted with a function of $\sin(angle)$. The direction of the minimum of this fit then was used to determine the best orientation that describes all 12 LEDs, and thus, the DOM that contains them, with on average around 1 degree (RMS) uncertainty. The orientations of the DOM and LEDs within, so found, were then used in all of the following simulations of the single LED flasher data set. Because the supporting (weight-bearing and data) cable was always (nominally during the deployment) routed between LEDs 11 and 12, this fit is also used to simulate the cable position with this same 1 degree uncertainty, plus some unknown (but thought to be better than 15 degrees) uncertainty in cable routing during the deployment.

\begin{figure}[h]
    \centering
    \includegraphics[width=0.31\textwidth]{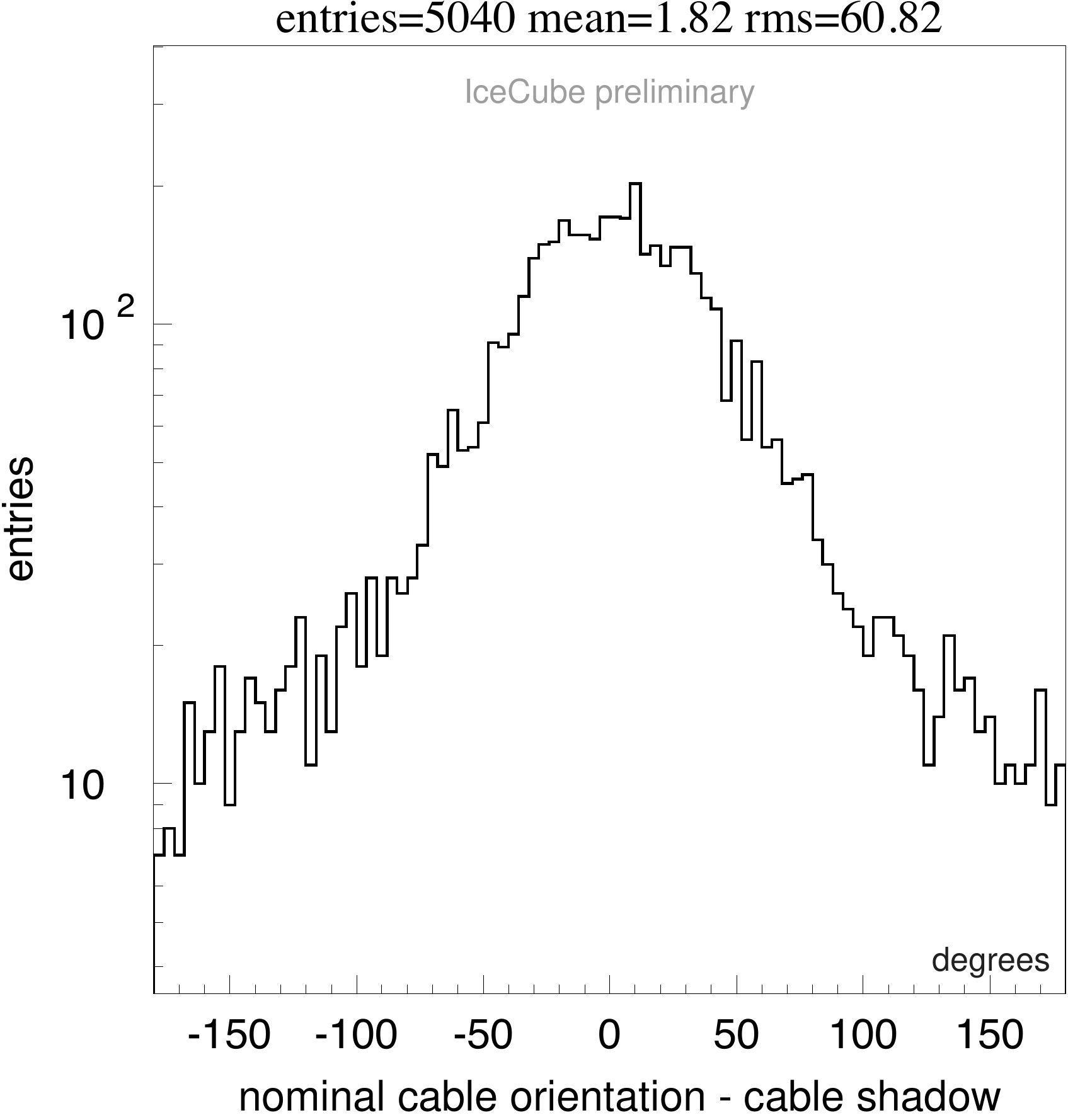}
    \hspace{1cm}
    \includegraphics[width=0.33\textwidth]{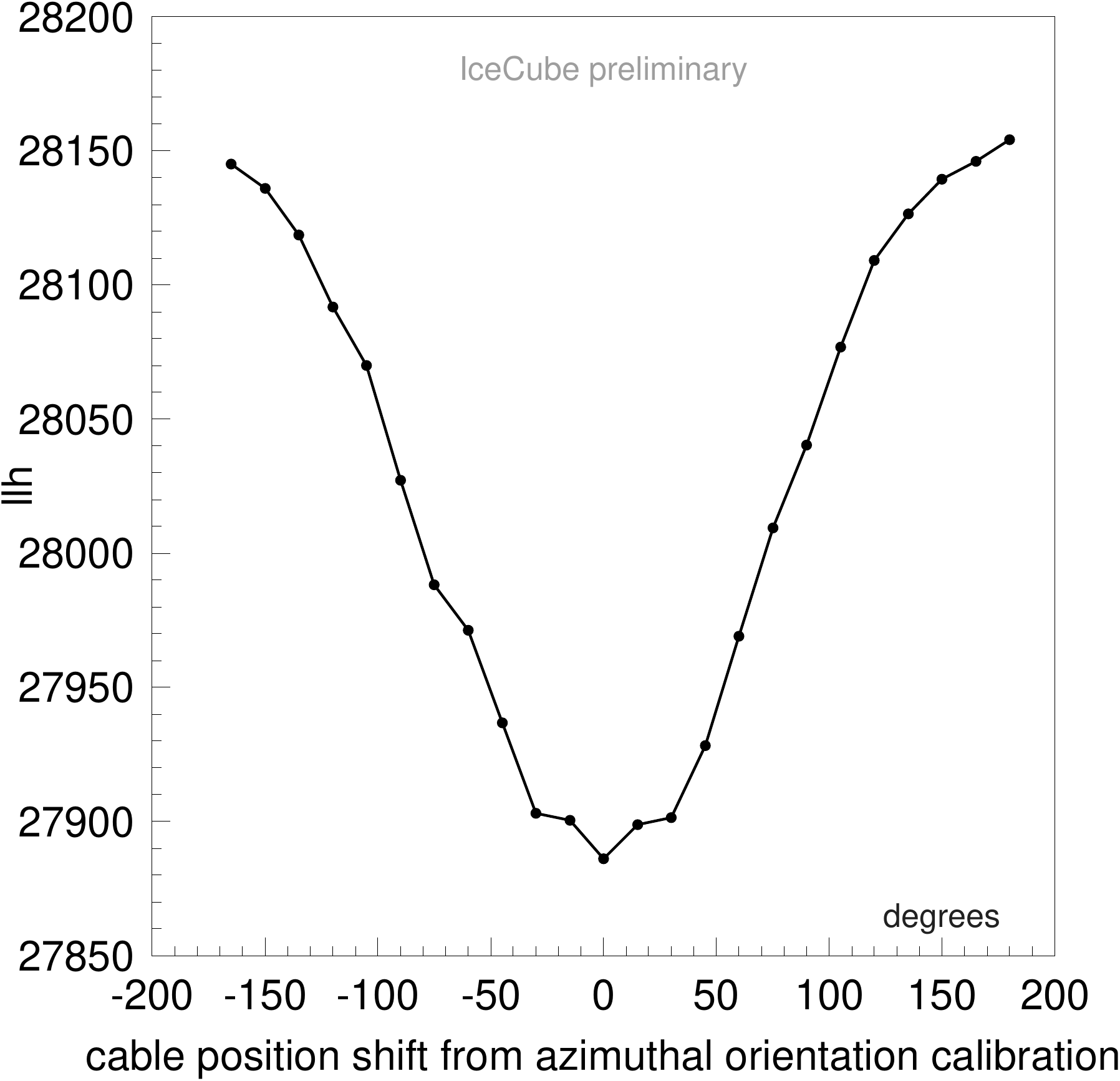}
    \caption{Left: histogram of differences (in degrees) between nominal cable position (with DOM orientation from single-LED orientation study) and reconstructed cable position from cable shadow study. Right: variation of {\it llh} sum over the entire detector as the cable position is shifted from nominal orientation.}
    \label{cable}
\end{figure}

We had also fitted the cable positions for all DOMs earlier with the {\it all-purpose} flasher data set by simulating the cable shadow on the receiving DOMs. The cable shadow was simulated in 1 degree increments (so, 360 possible locations for each DOM) and the best direction was chosen as the reconstructed cable position (determined by the best running average of {\it llh} values over the span of +-15 degrees). Since the cable is pressed against the DOM's surface, the distance between the cable and DOM is known and fixed, leaving azimuthal orientation the only fitted variable for each DOM. We have since correlated the cable position so determined with that from the {\it single-LED} orientation study described above and found that the shadow method is accurate to around 60 degrees RMS (assuming that the cable kept to within 15 degrees of its nominally specified position after deployment). This is shown in figure \ref{cable}. The figure also shows the variation in {\it llh} sum over the entire detector for the {\it single-LED} set, as the cable position determined with that set is shifted by uniform amount across all DOMs. The best description is achieved at zero shift from the reconstructed cable positions, as witnessed in the figure.

\section{Hole ice description and fits}
A set of two camera modules was installed at the bottom of the last string deployed in IceCube. The cameras are able to point at each other (up/down the hole) and captured a fascinating sequence of events that was the hole freezing over \cite{camera}. At the end of the freezing process they captured images that are now the basis of a model of the hole ice that is used in the discussion below. What the images showed is that most of the refrozen ice is extremely clear, possibly more clear than the original (warmed) ice just outside the drilled hole. All of the bubbles and other impurities were pushed in towards the center of the hole and frozen into a "bubbly column" that appears to be highly scattering column of ice that lights up when hit by the laser light not unlike the long fluorescent bulb when hit by light from a laser pointer. The rough geometrical layout observed is as follows: the central bubbly column of the hole ice has a diameter of about 1/2 of the diameter of a DOM. The refrozen hole has a diameter that is 1.5 that of a DOM. The DOM may lie anywhere within the refrozen hole, but what was observed was that both camera modules (same size as DOMs) were touching the wall of the refrozen ice on one side, with the bubbly central column covering a quarter of the DOM on the opposite side. This is the model that we accept in the following discussion, main parameters to be fit being the scattering coefficient of the bubbly central column, and the position of the DOM within the refrozen hole.

As mentioned above and seen in Figure \ref{azimuth}, in most DOMs all but a couple of the azimuthal {\it single-LED} {\it llh} curves look similar and have a pronounced peak and valley. The two curves that don't look like the rest are flatter and usually come in a matching pair (or two) of horizontal/tilted LEDs. Assuming the hole ice model described above, we can hypothesize that the two LEDs happen to fall inside the bubbly column of the hole ice, which scatters and randomizes the directions of outgoing photons. Because of that, no matter what azimuthal direction these two LEDs are simulated with, the resulting pattern will not match that in data, yielding a uniformly poor description across all azimuthal choices. We can usually identify such a "problem" region in most of the azimuthal orientation scans (and it is clearly seen in figure \ref{azimuth}).

\begin{figure}[h]
    \centering
    \begin{minipage}{.33\textwidth}
    \includegraphics[width=\linewidth]{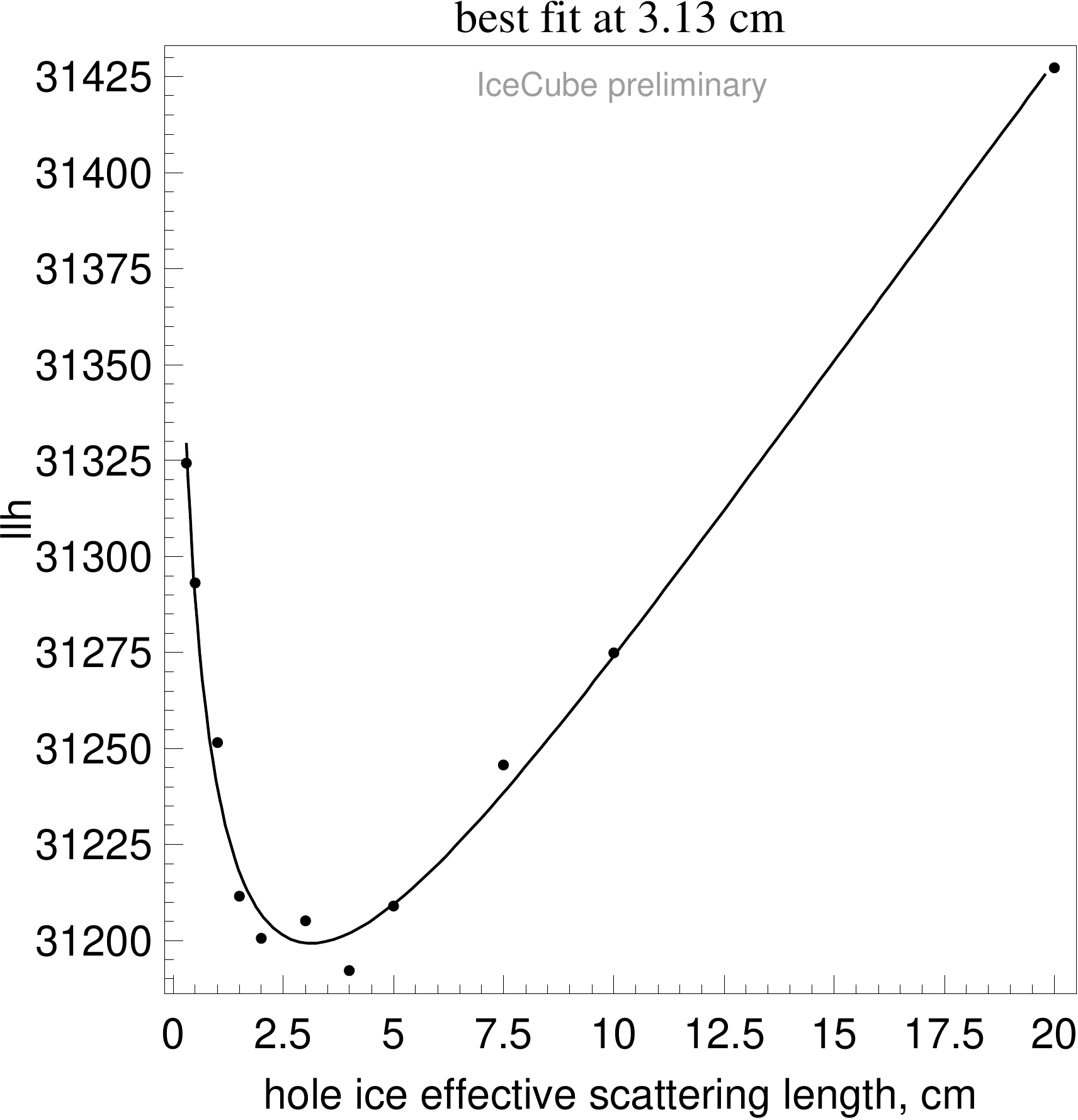}
    \end{minipage} \hfill
    \begin{minipage}{.33\textwidth}
    \includegraphics[width=\linewidth]{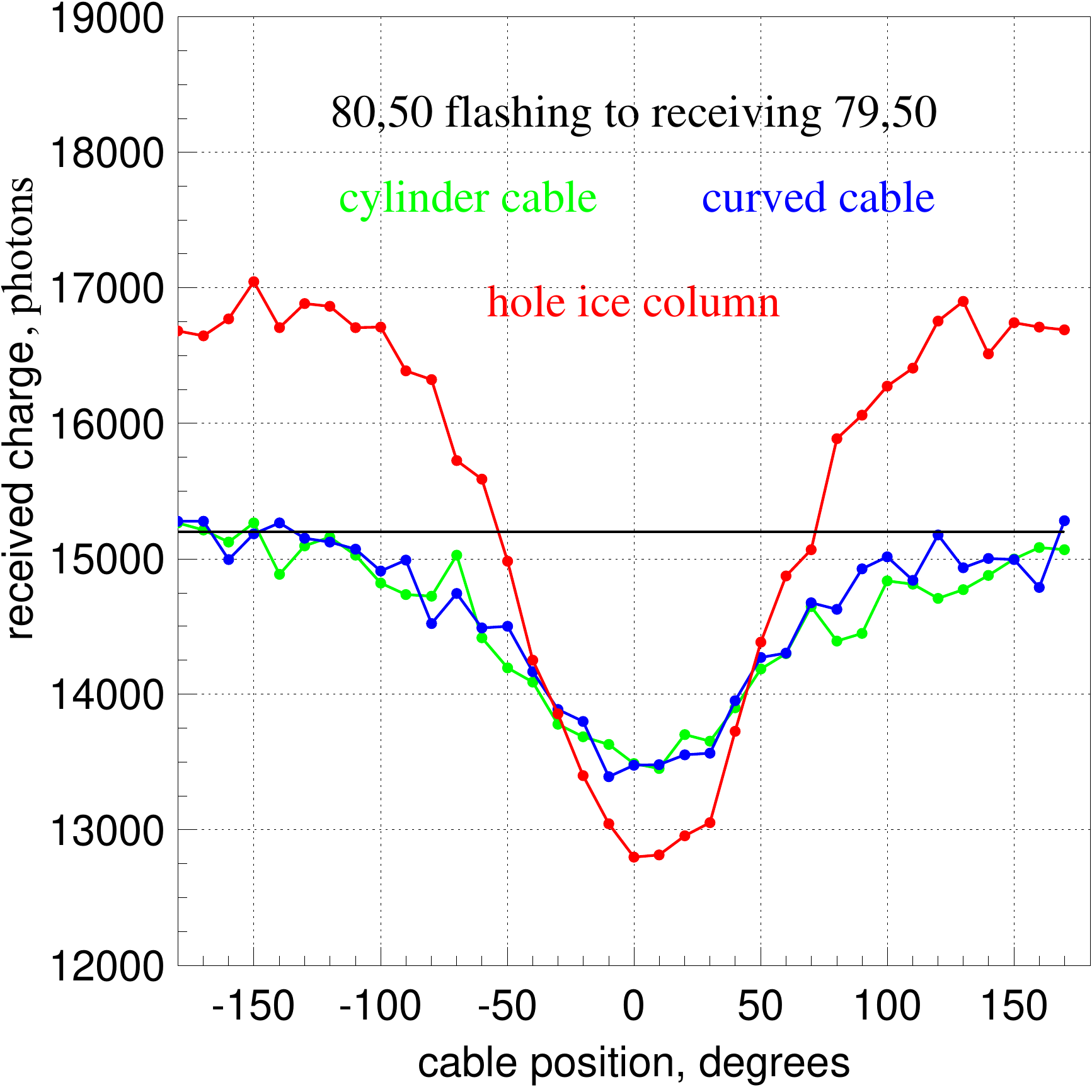}
    \end{minipage} \hfill
    \addtocounter{figure}{2}
    \begin{minipage}{.30\textwidth}
    \includegraphics[width=\linewidth]{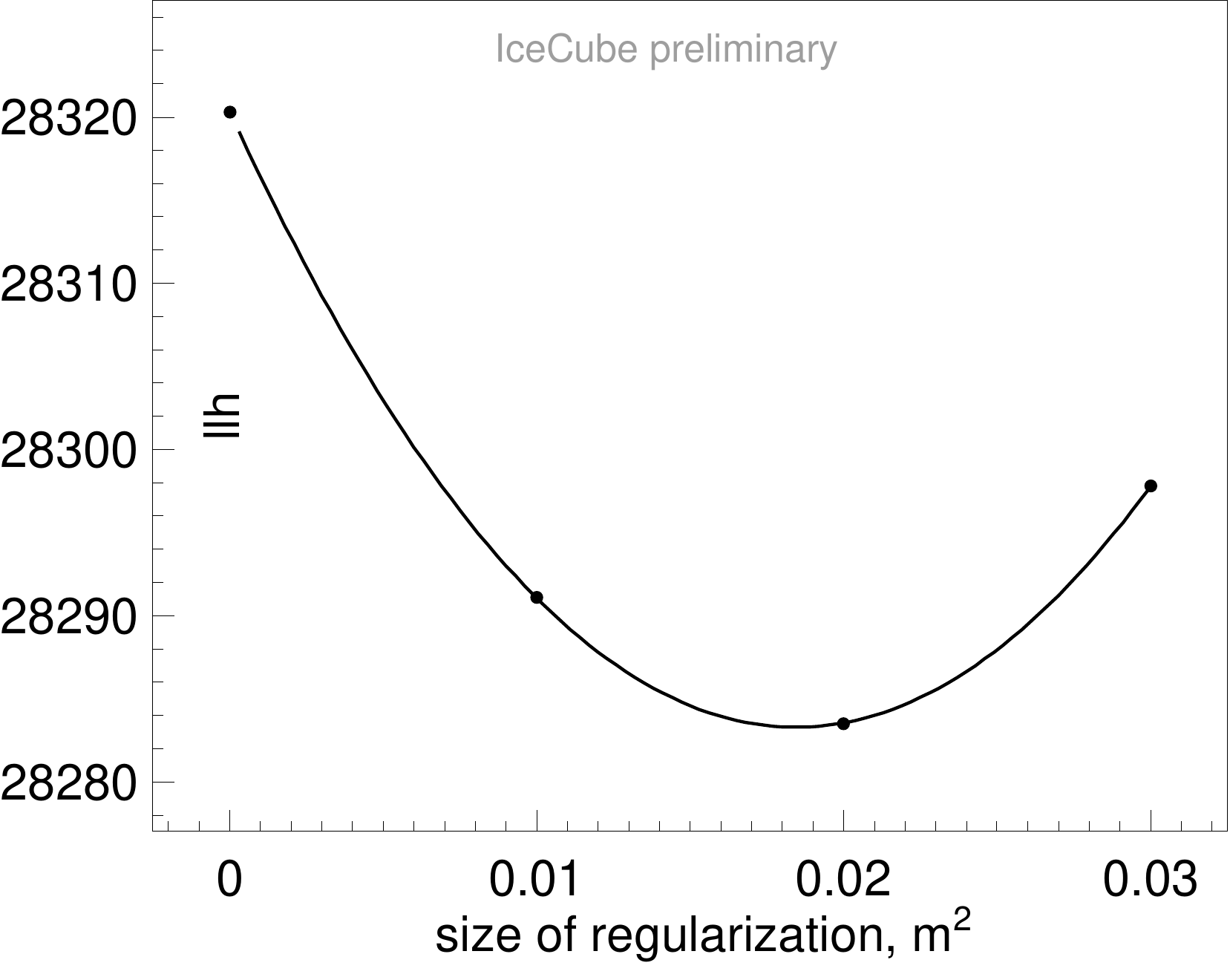}
    \caption{Fit to the best regularization parameter $a^2$.}
    \label{shadow2}
    \end{minipage}
    \addtocounter{figure}{-3}
    \begin{minipage}{.30\textwidth}
    \caption{{\it llh} scan over effective scattering length in the bubbly column of the hole ice.}
    \label{shadow1}
    \end{minipage} \hfill
    \begin{minipage}{.67\textwidth}
    \caption{Deficit/excess of observed number of photons due to cable shadow (cable simulated as straight cylinder and as a more realistic curve that bends around the DOM) and bubbly column of the hole ice.}
    \label{shadow3}
    \end{minipage}
    \addtocounter{figure}{1}
\end{figure}

We have since developed a method that relies on the direct simulation of the bubbly column of the hole ice. Before proceeding any further we need to know a rough estimate of the effective scattering length (distance at which the photon direction changes by an average of around one radian due to scattering) in the bubbly column. We centered the column (with a diameter of half the DOM diameter) on the centers of the DOMs for each string and simulated a range of scattering lengths. This is what we call a {\it receiver-side} fit in the following, as the simulated light from the LEDs misses the column of the bubbly ice (as they are just outside the column in this  simulated configuration), and it mainly just affects acceptance of the receiving DOM. We enforce that the bubbly column affects the photons only in the vicinity of the receiving DOM by only simulating its effect on photons with age greater than 20 ns. The column blocks and reflects photons, randomizing their directions where they are able to escape it. Figure \ref{shadow1} shows that the best description was achieved with an effective scattering length of 3\ cm.

As shown in \cite{dard} there is a significant correlation between the fitted width of the bubbly column and the effective scattering length of ice within (such that it roughly preserves the total amount of scattering centers in the bubbly column). We therefore fixed the diameter of the bubbly column to half the diameter of a DOM, as described above, and only fitted the effective scattering length to the data. The fitted effective scattering length varies only slightly with depth, and the average value of 3\ cm describes the entire depth range well.

\begin{figure}[h]
    \centering
    \begin{minipage}{0.85\textwidth}
    \includegraphics[width=0.245\linewidth,page=1]{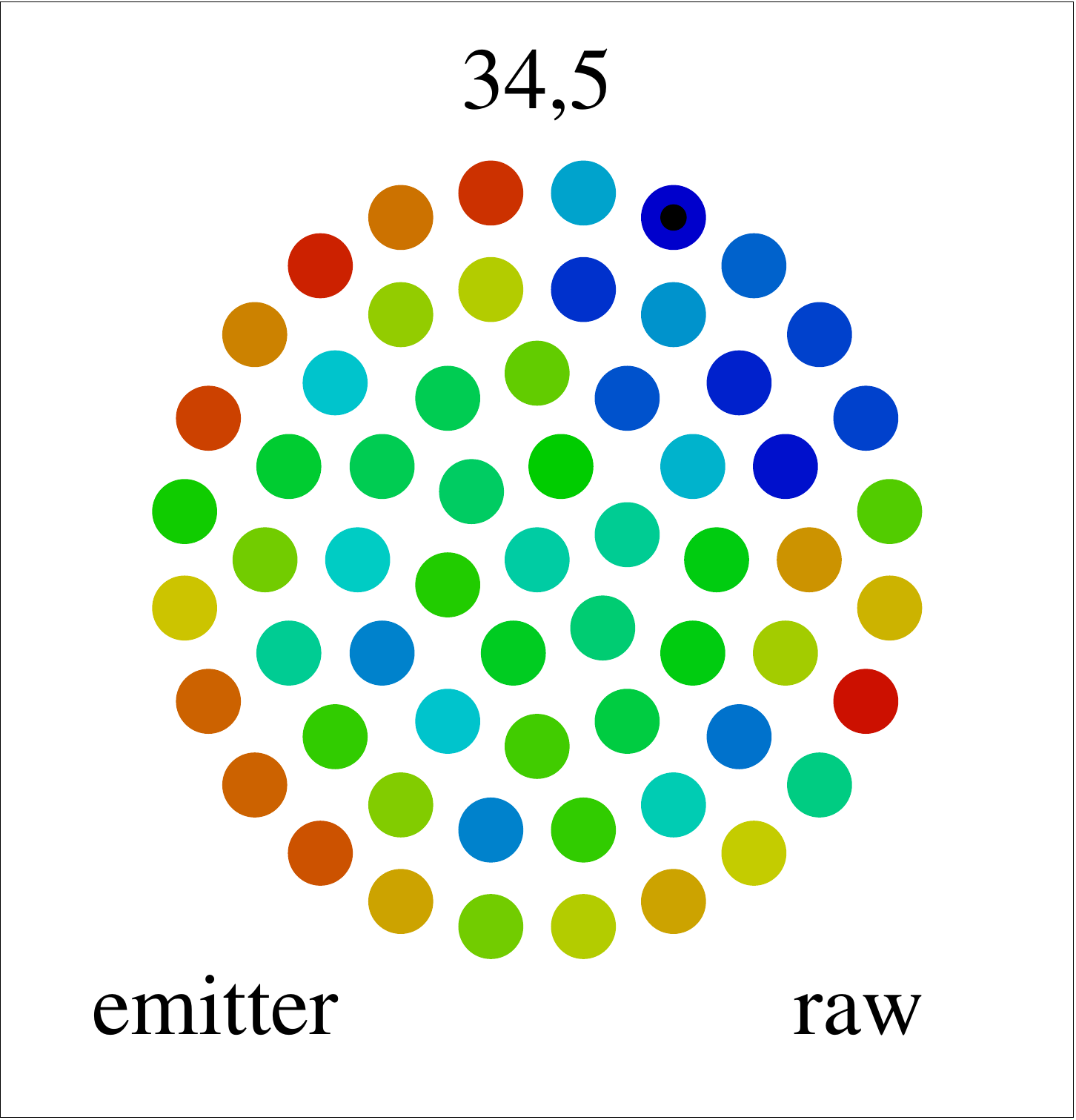}
    \includegraphics[width=0.245\linewidth,page=5]{figs/HI_scans.pdf}
    \includegraphics[width=0.245\linewidth,page=9]{figs/HI_scans.pdf}
    \includegraphics[width=0.245\linewidth,page=10]{figs/HI_scans.pdf}
    \includegraphics[width=0.245\linewidth,page=3]{figs/HI_scans.pdf}
    \includegraphics[width=0.245\linewidth,page=7]{figs/HI_scans.pdf}
    \includegraphics[width=0.245\linewidth,page=11]{figs/HI_scans.pdf}
    \includegraphics[width=0.245\linewidth,page=12]{figs/HI_scans.pdf}
    \end{minipage}
    \begin{minipage}{0.04\textwidth}
    \includegraphics[width=0.4\linewidth]{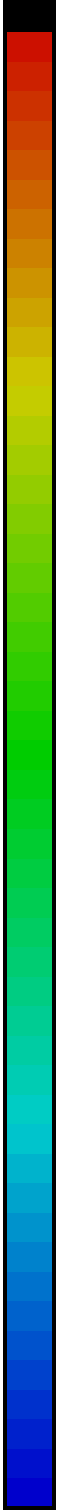}
    \end{minipage}
    \caption{Hole ice scan for DOM 34,5 (DOM 5 on string 34, top) and DOM 74,33 (bottom). From left to right: emitter-side scan, receiver-side scan, combined scan (sum of previous two), regularized (smoothed) scan. Points are shown in color sorted by their {\it llh} values, from blue (best) to red (worst). x and y axes show relative coordinates from DOM center to hole ice column center.}
    \label{scan}
\end{figure}

We have performed a fit to the relative position of all DOMs wrt.\ the bubbly column of the hole ice. The fit is performed in two stages: on the {\it receiver-side}, and on the {\it emitter-side}, where simulation of the bubbly column of the hole ice was only performed for photons younger than 20 ns from the emission time. In both cases we varied the relative position of the DOM wrt.\ the column on a grid of roughly uniformly spaced 61 points on a circle, either on the receiver or emitter side. Contributions from all receiving DOMs were combined into an {\it llh} sum for each of the {\it emitter-side} grid points, whereas all LEDs from all DOMs that contributed light to a receiving DOM were combined into an {\it llh} sum for each of the {\it receiver-side} grid points. Two examples of these scans are shown in figure \ref{scan}, where all points are sorted according to their {\it llh} value and the order is shown with color (from red/worst {\it llh} to blue/best {\it llh}). The two scans were then added together and a cluster/smoothing algorithm was applied as follows. The {\it llh} value at each point was averaged over the entire grid with weights equal to $1/(r^2+a^2)$, where $r$ is the distance to the weighted point and $a^2$ is a {\it regularization} parameter, which was optimized to produce the set of relative DOM to hole ice column positions which resulted in the best description of the entire {\it single-LED} data set. A value of $a^2=0.02$\ m$^2$ was so found (shown in figure \ref{shadow2}). The smoothing procedure described here improved most of the scans that looked ambiguous due to statistical fluctuations to the point where the choice of the best point was visually compelling and thus deemed robust (as is evident from the second example in figure \ref{scan}).

\begin{figure}[h]
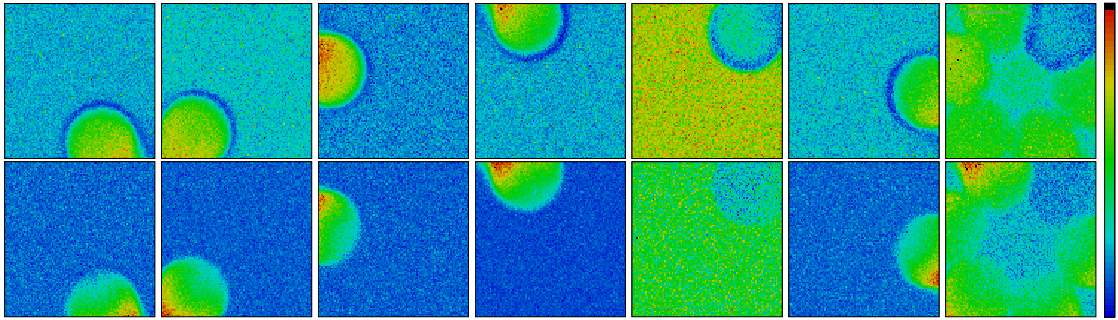

    \centering
    \begin{minipage}{0.95\textwidth}
    \includegraphics[width=0.138\linewidth,page=4]{figs/HI_hires.pdf}
    \includegraphics[width=0.138\linewidth,page=5]{figs/HI_hires.pdf}
    \includegraphics[width=0.138\linewidth,page=6]{figs/HI_hires.pdf}
    \includegraphics[width=0.138\linewidth,page=7]{figs/HI_hires.pdf}
    \includegraphics[width=0.138\linewidth,page=8]{figs/HI_hires.pdf}
    \includegraphics[width=0.138\linewidth,page=9]{figs/HI_hires.pdf}
    \includegraphics[width=0.138\linewidth,page=2]{figs/HI_hires.pdf}
    \includegraphics[width=0.138\linewidth,page=10]{figs/HI_hires.pdf}
    \includegraphics[width=0.138\linewidth,page=11]{figs/HI_hires.pdf}
    \includegraphics[width=0.138\linewidth,page=12]{figs/HI_hires.pdf}
    \includegraphics[width=0.138\linewidth,page=13]{figs/HI_hires.pdf}
    \includegraphics[width=0.138\linewidth,page=14]{figs/HI_hires.pdf}
    \includegraphics[width=0.138\linewidth,page=15]{figs/HI_hires.pdf}
    \includegraphics[width=0.138\linewidth,page=3]{figs/HI_hires.pdf}
    \end{minipage}
    \begin{minipage}{0.04\textwidth}
    \includegraphics[width=0.27\linewidth]{figs/color-cropped.pdf}
    \end{minipage}
    \caption{Fine-grid hole ice scan for DOM 34,5: tilted LEDs 1-6 (top) and horizontal LEDs 7-12 (bottom). From left to right: 6 individual LEDs followed by their sum. Color scale is from blue (best) to red (worst). x and y axes show relative coordinates from DOM center to hole ice column center.}
    \label{finescan}
\end{figure}

We have also performed high-resolution scans on a fine grid of 101x101 points for the two examples discussed above, with one of them shown in figure \ref{finescan}. Only {\it emitter-side} scans are shown because they are much more visually appealing and contain interesting sharp features unlike the {\it receiver-side} scans. Both the scans for each individual LED and the sum totals for tilted and horizontal LEDs are shown. In this example, a matched pair of the LEDs (5 and 11) appear to shine directly into the bubbly column of the hole ice. Only for these two LEDs the description of data with simulation improves when the bubbly column is placed in front of them. For all other LEDs the {\it llh} value is best and roughly the same when the bubbly column is anywhere but in front of them, and degrades significantly when it is in front and blocking the emitted LED light.

In order to visualize the effect of the hole ice column we simulated positioning it at different angles around a DOM with the distance between the DOM center and the bubbly column center of 1/2 of DOM radius (this is the scenario where the DOM touched the wall of the refrozen hole). This, together with a somewhat similar effect from the cable shadow, is shown in figure \ref{shadow3}. Both the bubbly column and cable block photons coming from the side of the DOM that they are covering, but the bubbly column also enhances the number of detected photons when the column is on the opposite side. This we attribute to a {\it reflection} effect, where the photons enter the column and are redirected towards the DOM.

\section{Effective description of local ice, DOM tilt, and relative in-ice sensitivities}
As discussed in \cite{spice}, in order to accelerate the photon propagation simulation we {\it oversize} the DOM by making it larger in the directions perpendicular to the incoming photon (usually by a factor of 5, making it into a sort of a "pancake"). This increases the cross section by a factor of 25, thus reducing the number of photons that need to be propagated by this same factor, resulting in very substantial computational savings. Such a procedure, however, results in an inability to simulate the local DOM effects, such as hole ice, precisely. We also usually switch to an {\it effective} angular DOM sensitivity description, where the DOM acceptance is modeled as a function of incoming photon direction (a plane wave treatment), instead of a more accurate approach where the acceptance depends on photon landing coordinates on the DOM surface (so that the photon is accepted only when it enters the DOM sphere near the PMT cathode). The {\it effective} plane wave approach allows one to modify the angular sensitivity curve measured in the lab to account for the local (mainly {\it hole ice}) effects as accurately as possible (when {\it oversizing}). This lowers acceptance somewhat for photons arriving with the direction into the PMT and raises acceptance for photons arriving from the opposite direction. This approach allows us to continue using the {\it oversized} DOMs in the simulation while taking into account some of the effect of the bubbly column of the hole ice.

We can improve this further by converting the relative positioning of the bubbly column wrt.\ the DOM into a DOM {\it tilt}. Tilting a DOM to one side is implemented by applying the chosen DOM acceptance model as a function of the photon direction wrt.\ the tilted DOM axis rather than the nominally vertical direction. It results in an effect similar to that shown in figure \ref{shadow3}: DOM acceptance increases on one side and drops on the other. Because the cable shadow may also look like a similar effect, the cable was placed and fixed at the nominal position as found from the azimuthal orientation study. We have fitted the DOM tilts for all DOMs in the detector by simulating the DOM directions on a icosahedron-generated grid of 642 points. The best 10 directions were averaged to produce the reconstructed {\it tilt} direction. While we started the tilt study as an attempt to improve the effective description of local ice when simulating {\it oversized} DOMs, we had also repeated the fit with the nominal-size DOMs and with the full simulation of the bubbly column of the hole ice with effective scattering and relative wrt.\ DOM positioning reconstructed as described above. We found a distribution of DOM tilts that is very similar, which is puzzling and is a subject of an on-going study. Curiously, we compared the tilts found here to the measurement with the built-in inclinometer that is installed in 48 of the in-ice DOMs, and found a correlation between the two outliers in the inclinometer data with the tilts found with the method discussed here.

Finally we also fitted the overall relative in-ice DOM sensitivities (wrt.\ their nominal average baseline value) to the {\it single-LED} data set. The resulting values form a distribution with a width of around 8.6\% (RMS). We have also split the fit using only subsets of LEDs, e.g., only tilted or only horizontal LEDs. The resulting shifts from the fit using all LEDs are within 1\% of each other, which demonstrates good stability of the result. We note here that the relative in-ice DOM sensitivity may not necessarily match the relative DOM efficiency as measured in the lab due to local effects such as positioning of the DOM within the hole and relative to the cable and bubbly column, as well as due to possible deposits of dust and other debris on the DOM surface as observed in the pictures taken by the two in-ice cameras. In fact, the correlation between relative in-ice sensitivities fitted here and lab-measured relative DOM efficiencies is virtually non-existent.

\section{Discussion of fit results}
All of the fits described here had employed the same likelihood construction \cite{llh} as in most other recent ice work, e.g., in \cite{bfr,spice}. The minus log likelihood, denoted in this report as {\it llh}, is akin to the saturated Poisson likelihood, and can similarly be used as a measure of the goodness-of-fit. Here the {\it llh} value is a sum over around 60000 LEDs (12 per DOM, around 5000 DOMs), and has a statistical and numerical uncertainty estimated at around 12. Built into this likelihood construction is an assumption that the data and simulation at the best fit point (in the limit of infinite statistics) disagree at a configurable level of around 10\%. We call this disagreement {\it model error} and can gauge it after the fit by estimating the width of the distribution of ratios of charges in data and simulation. The effects fitted and described in this report improve both the {\it llh} and {\it model error} in our description of the LED calibration data. Although the estimated {\it llh} and {\it model error} values depend strongly on the specific simulation setup used in data/simulation comparison (such as amount of statistics simulated, and range of received charge used in the comparison, we can fix these and only vary the underlying ice/detector model. We start with {\it llh}=28472, {\it model error=}14.1\% (for the ice model described in \cite{bfr}). Adding the full hole ice description as calibrated here these values are reduced to 28307 and 13.6\%. Further changing to the in-ice DOM sensitivities fitted here reduces these values to 27892 and 10.0\%. Finally adding tilt we get 27644 and 9.9\%. The numbers are given with {\it oversized} DOMs (factor of 5). Simulation with nominal-size DOMs results in {\it llh}=27542 and an estimated {\it model error} of 9.8\%. We have tested the effects fitted here on real-life IceCube events (similar to figure 7 in \cite{bfr}) and witnessed an improvement in the description of those events.

\bibliographystyle{ICRC}
\bibliography{main}

% Full authors list (ONLY FOR COLLABORATIONS)
\clearpage
\section*{Full Author List: IceCube Collaboration}

% \noindent \textbf{Note comment afterwards:} Collaborations have the possibility to provide an authors list in xml format which will be used while generating the DOI entries making the full authors list searchable in databases like Inspire HEP. For instructions please go to icrc2021.desy.de/proceedings or contact us under icrc2021proc@desy.de.\\

% \scriptsize
% \noindent
% first.author$^1$, 
% second.author$^2$, 
% third.author$^3$ % .... more names
% and 
% last.author$^{n}$ \\

% \noindent
% $^1$first.affiliation.
% $^2$second.affiliation. % .... more affiliation
% $^{m}$last.affiliation.

\scriptsize
\noindent
R. Abbasi$^{17}$,
M. Ackermann$^{59}$,
J. Adams$^{18}$,
J. A. Aguilar$^{12}$,
M. Ahlers$^{22}$,
M. Ahrens$^{50}$,
C. Alispach$^{28}$,
A. A. Alves Jr.$^{31}$,
N. M. Amin$^{42}$,
R. An$^{14}$
% ...
\\

\noindent
% ...
$^{12}$ Universit{\'e} Libre de Bruxelles, Science Faculty CP230, B-1050 Brussels, Belgium \\
% ...
$^{14}$ Department of Physics and Laboratory for Particle Physics and Cosmology, Harvard University, Cambridge, MA 02138, USA \\
% ...
$^{17}$ Department of Physics, Loyola University Chicago, Chicago, IL 60660, USA \\
$^{18}$ Dept. of Physics and Astronomy, University of Canterbury, Private Bag 4800, Christchurch, New Zealand \\
% ...
$^{22}$ Niels Bohr Institute, University of Copenhagen, DK-2100 Copenhagen, Denmark \\
% ...
$^{28}$ D{\'e}partement de physique nucl{\'e}aire et corpusculaire, Universit{\'e} de Gen{\`e}ve, CH-1211 Gen{\`e}ve, Switzerland \\
% ...
$^{31}$ Karlsruhe Institute of Technology, Institute for Astroparticle Physics, D-76021 Karlsruhe, Germany  \\
% ...
$^{42}$ Bartol Research Institute and Dept. of Physics and Astronomy, University of Delaware, Newark, DE 19716, USA \\
% ...
$^{50}$ Oskar Klein Centre and Dept. of Physics, Stockholm University, SE-10691 Stockholm, Sweden \\
% ...
$^{59}$ DESY, D-15738 Zeuthen, Germany \\
% ...

\subsection*{Acknowledgements}

\noindent
USA {\textendash} U.S. National Science Foundation-Office of Polar Programs,
U.S. National Science Foundation-Physics Division,
U.S. National Science Foundation-EPSCoR,
Wisconsin Alumni Research Foundation,
Center for High Throughput Computing (CHTC) at the University of Wisconsin{\textendash}Madison,
Open Science Grid (OSG),
Extreme Science and Engineering Discovery Environment (XSEDE),
Frontera computing project at the Texas Advanced Computing Center,
U.S. Department of Energy-National Energy Research Scientific Computing Center,
Particle astrophysics research computing center at the University of Maryland,
Institute for Cyber-Enabled Research at Michigan State University,
and Astroparticle physics computational facility at Marquette University;
Belgium {\textendash} Funds for Scientific Research (FRS-FNRS and FWO),
FWO Odysseus and Big Science programmes,
and Belgian Federal Science Policy Office (Belspo);
Germany {\textendash} Bundesministerium f{\"u}r Bildung und Forschung (BMBF),
Deutsche Forschungsgemeinschaft (DFG),
Helmholtz Alliance for Astroparticle Physics (HAP),
Initiative and Networking Fund of the Helmholtz Association,
Deutsches Elektronen Synchrotron (DESY),
and High Performance Computing cluster of the RWTH Aachen;
Sweden {\textendash} Swedish Research Council,
Swedish Polar Research Secretariat,
Swedish National Infrastructure for Computing (SNIC),
and Knut and Alice Wallenberg Foundation;
Australia {\textendash} Australian Research Council;
Canada {\textendash} Natural Sciences and Engineering Research Council of Canada,
Calcul Qu{\'e}bec, Compute Ontario, Canada Foundation for Innovation, WestGrid, and Compute Canada;
Denmark {\textendash} Villum Fonden and Carlsberg Foundation;
New Zealand {\textendash} Marsden Fund;
Japan {\textendash} Japan Society for Promotion of Science (JSPS)
and Institute for Global Prominent Research (IGPR) of Chiba University;
Korea {\textendash} National Research Foundation of Korea (NRF);
Switzerland {\textendash} Swiss National Science Foundation (SNSF);
United Kingdom {\textendash} Department of Physics, University of Oxford.

\end{document}